\documentclass[10pt,aps,physrev,groupedaddress]{revtex4-2}

\usepackage{amsmath}
\usepackage[]{hyperref}
\hypersetup{
    colorlinks=true,      
    linkcolor=blue,       
    citecolor=blue,      
    urlcolor=cyan,        
    filecolor=magenta     
}
\usepackage{orcidlink}

\usepackage{graphicx}
\usepackage{caption}
\usepackage[normalem]{ulem}

\usepackage{amssymb}
\usepackage{mathabx}
\usepackage{tikz}
\definecolor{cb2green}{RGB}{77,175,74}
\definecolor{cb2blue}{RGB}{55,126,184}
\definecolor{cb2red}{RGB}{228,26,28}

\newcommand{\legline}[2]{
  \protect\tikz[baseline=-0.6ex]{%
    \protect\draw[#1, thick] (0,0) -- (0.6cm,0);
    \protect\node[#1] at (0.3cm,0) {\scalebox{0.8}{$#2$}};
  }%
}

\begin{document}


\title{Coherent structure modulation and recovery in drag-reduced turbulent boundary layers}


\author{Traian Bistriceanu}
\altaffiliation{These authors contributed equally to this work.}

\author{Vyacheslav A. Kovalyov}
\altaffiliation{These authors contributed equally to this work.}

\author{Pradyumn Mishra}

\author{Roann A. A. Kerbergen}

\author{Max W. Knoop\orcidlink{0009-0008-8848-3006}}
\email[]{Contact Author: m.w.knoop@tudelft.nl}
\author{Bas W. van Oudheusden\orcidlink{0000-0002-7255-0867}}

\affiliation{Faculty of Aerospace Engineering, Delft University of Technology, 2629HS Delft, The Netherlands}


\date{\today}

\begin{abstract}


This study investigates the coherent structures in drag-reduced turbulent boundary layers.
The motivation is the direct link between turbulent fluctuations and skin-friction drag via the Reynolds shear stress and its associated coherent structures.
Drag reduction is achieved by a steady square-wave forcing of the spanwise wall velocity, based on the experiments in \citeauthor{knoop2025response} (\emph{Phys. Rev. Fluids}, 10, 2025).
Particle tracking velocimetry data are analyzed for a non-actuated reference case and actuation at forcing amplitude $A^+ = 12$ ($+$ denotes viscous scaling) for three streamwise forcing wavelengths, which correspond to sub-optimal ($\Lambda_x^+ = 471$), near-optimal ($\Lambda_x^+ = 942$), and post-optimal ($\Lambda_x^+ = 1884$) drag reduction conditions.
Conditionally averaged fields on large-scale bursts of turbulent kinetic energy show that forcing suppresses near-wall ejections across all cases, while outer-layer sweep suppression strengthens with $\Lambda_x^+$.
Post-optimal forcing exhibits streamwise-periodic attenuation and recovery of turbulence.
The recovery phenomenon is caused by an enhancement of very small scales, significantly smaller than those typically energetic in wall turbulence, and is linked to the emergence of small-scale bursting events. 
While these small-scale bursts are statistically insignificant in the non-actuated and sub-optimal cases, their frequency increases by a factor of four between the near-optimal and post-optimal cases. 
The small-scale bursts exhibit uniform signatures and intensities, hinting at the possible universality of the recovery phenomenon.
A wavelet analysis shows that, in the post-optimal case, very small scales increase progressively in the streamwise direction in regions where the wall velocity remains constant, driving a cyclic pattern of small-scale re-energization and suppression consistent with earlier statistical analysis.
This turbulence modulation mechanism is scale-selective: while small-scale structures emerge periodically, large-scale motions are more effectively suppressed as the forcing wavelength increases.

\end{abstract}


\maketitle


\section{Introduction \label{sec: intro}}
Turbulent skin-friction drag is responsible for approximately 50\% of the energy losses in the aerospace industry \citep{Ricco2021}, and can reach up to 90\% in other sectors, such as long-distance oil pipelines.
Flow control aimed at turbulent drag reduction (DR) can therefore yield substantial energy savings.
Passive control strategies achieve drag reduction through static surface modifications, such as (sinusoidal) riblets \citep{Peet2008, cafiero2024manipulation}, dimpled surfaces \citep{vanNesselrooij2016}, and wavy-wall geometries \citep{Ghebali2017, Chernyshenko2013part1}. 
Active strategies include wall-normal suction/blowing \citep{Kametani2011, Kametani2015, Kornilov2012}, opposition control \citep{Choi1994, dacome2024opposition}, wall cooling \citep{Xin2023}, and near-wall spanwise forcing \citep{Jung1992SuppressionOscillations, Laadhari1994, Choi2002, Quadrio2009Streamwise-travellingReduction, Ricco2021}.
Among these, spanwise wall-forcing stands out for its high DR efficacy, yielding values up to $40\%$ for the oscillating wall (OW) (i.e., purely temporal) approach \citep{Jung1992SuppressionOscillations, Laadhari1994}, and up to  $48\%$ for the streamwise-travelling wave (STW) (i.e., spatio-temporal) one \citep{Quadrio2009Streamwise-travellingReduction}.

Skin-friction drag is generated by the viscous no-slip condition and is proportional to the wall-normal gradient of mean-streamwise-velocity.
The skin-friction coefficient is defined as $C_f = \frac{2\nu}{U_\infty^2}(\frac{\partial\overline{U}}{\partial y})_{(y=0)}$, where $U_\infty$ is the free-stream velocity and $\nu$ is the kinematic viscosity of the fluid.
In this study, we consider a wall-bounded turbulent flow, represented with a coordinate system $(x,y,z)$ and velocity components $(U,V,W)$, corresponding to the streamwise, wall-normal, and spanwise directions. 
The velocity is decomposed into an ensemble (time) averaged mean and fluctuating component, indicated by an overbar and lowercase, respectively, i.e., $U= \overline{U}+u$. In a spanwise homogeneous zero-pressure-gradient (ZPG) turbulent boundary layer (TBL), skin-friction and turbulent fluctuations are related through the Reynolds shear stress ($-\overline{uv}$) in the (streamwise) mean-momentum balance \citep{Klewicki2010}:
    \begin{equation}
        \label{eq:U-momeq}
        \overline{U}\frac{\partial \overline{U}}{\partial x} + \overline{V}\frac{\partial \overline{U}}{\partial y} = \nu \frac{\partial^2  \overline{U}}{\partial y ^2} - \frac{\partial \overline{uv}}{\partial y}. 
    \end{equation}
A formal relationship between $C_f$ and $-\overline{uv}$ can be derived by wall-normal integration of \eqref{eq:U-momeq}, such as the \citet*{Fukagata2002} identity, which is strictly valid only for internal flows \citep{ricco2022integral}.
To extend this concept to TBL flows, \citet{Elnahhas2022} integrated the momentum deficit equation, wherein the turbulent contribution to $C_f$ reads:
    \begin{equation}
    \label{eq: Cf wall-normal integration}
          C_{f,turb}(x) = 2\int_0^\infty \frac{-\overline{uv}(x,y)}{U_\infty^2 l(x)}\mathrm{d}y,
    \end{equation}
where $l(x)$ is a reference length that must be introduced for a boundary layer  (e.g., the momentum thickness).
This link between $C_f$ and $-\overline{uv}$ in internal and external TBL flow motivates our investigation into the organization of drag-producing coherent structures associated with the Reynolds shear stress.

Interest in the coherent flow structures emerged with the near-wall flow visualizations of \citet{kline_structure_1967}, who first revealed the now well-established high- and low-speed streamwise streaks of $u$-fluctuations.
These observations led to the self-sustaining cycle (SSC) paradigm between near-wall streaks and quasi-streamwise vortices (QSV) \citep{Jimenez1999}.
Meanwhile, the joint probability density function between streamwise and wall-normal velocity fluctuations $P(u,v)$, formalized as quadrant analysis by \citet{Wallace1972} and \citet{Lu1973}, established the importance of the wall-normal velocity fluctuations in the redistribution of streamwise momentum.
The two most energetic quadrants are associated with motions classified as ejections (Q2) carrying low-momentum fluid ($u<0$) away from the surface ($v>0$), and sweeps (Q4) transporting high-momentum ($u>0$) towards the wall ($v<0$).
Given that the near-wall turbulence is dominated by ejection and sweep events, the anti-correlation between $u$ and $v$ causes the negative sign of the Reynolds shear stress, leading to a positive $C_f$ contribution per equation \eqref{eq: Cf wall-normal integration}.
Building on these concepts, \citet{Adrian2007HairpinTurbulence} consolidated the paradigm that hairpin-like coherent vortical structures populate the turbulent flow, which generate concentrated regions of Q2 and Q4 activity, significantly contributing to the Reynolds shear stress.
The near-wall flow visualizations of \citet{kline_structure_1967} also led to the concept of bursting, as intermittent regions of high turbulent kinetic energy (TKE) production \citep{Kim1971}. Subsequent studies attributed this process to the impingement of high-speed sweep motions onto low-speed ejections, which generate intense local shear layers and trigger the violent bursting of near-wall structures \citep{offen1975bursting}.
Bursts can be detected and visualized by conditional averaging. A detection procedure, variable-interval time-averaging (VITA), was initially introduced by \citet{Blackwelder1976} for single-point hot-wire measurements and later extended by \citet{Kim1985} to variable-interval space-averaging (VISA) as its spatial equivalent for flow-field data.

Spanwise wall forcing has been reported to lead to a strong suppression of turbulence and a weakening of its near-wall structures.
For example, a notable reduction of ejections and sweeps was reported by \citet{Choi2002} and \citet{Ricco2004}, and an attenuation of the streamwise vorticity of the QSVs by \citet{Choi2001, Choi2002, Kempaiah20203-dimensionalOscillation}, and \citet{Yakeno2014}.
Among others, these studies proposed that the transverse shear layer, referred to as the Stokes layer, effectively displaces the streaks from the overlying QSVs, thereby disrupting the SSC.
At near-optimal drag-reducing forcing conditions, \citet{Yakeno2014} (for the OW technique) and \citet{Gallorini2022} (extending the analysis for the STW case) showed that the largest $C_f$ reduction was caused by the suppression of ejections (Q2), followed by the suppression of sweeps (Q4).
At non-optimal or drag-increasing conditions, however, Q4 events further increased the skin-friction drag during certain phases of the actuation.

 While OW and STW type wall-forcing have been studied extensively, the purely spatial (steady) forcing introduced in \citet{Viotti2009StreamwiseReduction} has received limited attention, despite its high DR efficacy. Spatial forcing imposes a steady spanwise wall velocity that varies periodically in the streamwise direction. This strategy was investigated experimentally for the first time as a proof-of-concept in \citet{Knoop_2024} and later extended to a full-scale setup in \citet{knoop2025response}, spanning a streamwise extent of approximately $11.5$ inflow boundary-layer thicknesses ($\delta_0$), to establish the fully-developed forcing effects.
 The actuation surface in \citet{knoop2025response} comprises 48 spanwise running belts to impose a square-wave (SqW) type  spanwise wall velocity given by:
    \begin{equation}
        \label{W_w}
        W_w(x) = A\,\mathrm{sgn}\!\left[\sin\!\left(\frac{2\pi}{\Lambda_x}x\right)\right],
    \end{equation}
where $A$ is the forcing amplitude and $\Lambda_x$ is the forcing wavelength.
\citet{knoop2025response} established the streamwise evolution of a TBL flow at constant $A^+ = 12$ for three test cases by varying the wavelength in the sub-optimal ($\Lambda_x^+ = 471$), near-optimal ($\Lambda_x^+ = 942$), and post-optimal ($\Lambda_x^+ = 1884$) DR regimes. Here, the `$+$' superscript indicates viscous scaling, based on reference (i.e., non-actuated) conditions.
The DR and turbulence statistics were validated according to the established literature \citep{Viotti2009StreamwiseReduction,Gatti_2016} on sinusoidal forcing (i.e., omitting the $\mathrm{sgn}$ operator in \eqref{W_w}).
A key insight was the observed periodic attenuation and recovery of turbulence across the actuation phase in the large-wavelength post-optimal case. In \S\ref{sec:intro-recovery}, this intra-phase recovery effect (i.e., a phase-wise variation of statistics) is briefly reviewed.
While \citet{knoop2025response} focused on DR and the spatial evolution of statistics, the impact of SqW forcing on the coherent structures and their recovery remains an open question that is addressed in this study (refer to \S\ref{sec:intro-research approach}).

\begin{figure}[t]
    \centering
    \includegraphics[]{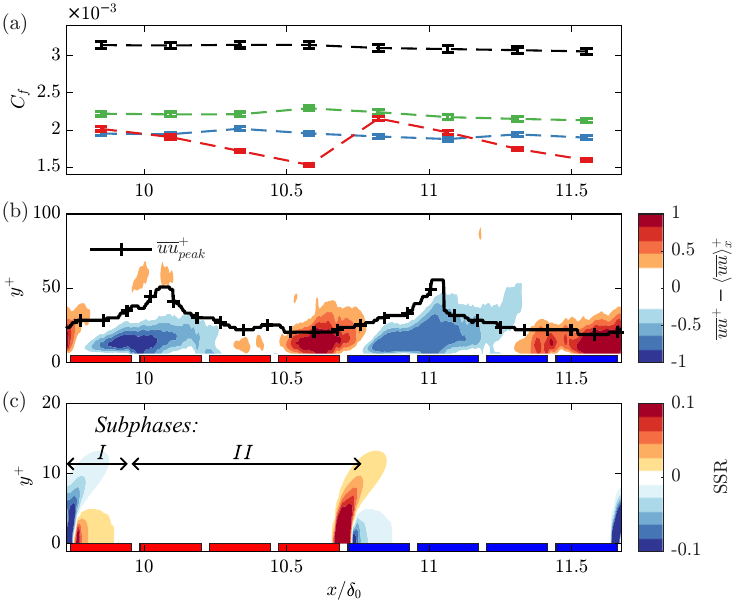}
    \caption{(a) Streamwise evolution of skin-friction coefficient ($C_f$) for the non-actuated case (black), sub-optimal (green; $\Lambda_x^+ = 471$), near-optimal (blue; $\Lambda_x^+ = 942$), and post-optimal (red; $\Lambda_x^+ = 1884$) forcing regimes; note that the markers correspond to its evaluation on the center of each belt; dashed connecting lines are added for interpretability only.
    For the post-optimal case $\Lambda_x^+ = 1884$:
    (b) $\overline{uu}^+ - \langle \overline{uu}\rangle^+_x$, i.e., the deviation of $\overline{uu}^+$ from its streamwise-averaged value, the black line indicates the $y$-location of the $\overline{uu}^+$ inner peak, and (c) analytical solution of SSR \citep{Knoop_2024}.
    The shaded red (blue) patches in (b,c) denote the individual belts and their positive (negative) spanwise wall velocity $W_w$, and $x=0$ corresponds to the start of the actuation surface. 
    Reproduced from \citet{knoop2026internal} and based on experimental results in \citet{knoop2025response}.}
    \label{fig:intro-recovery}
\end{figure}

\subsection{Recovery effects for post-optimal forcing conditions \label{sec:intro-recovery}}
The intra-phase recovery effect has been observed previously in literature for long-period OW forcing in \citet{Touber2012},  and \citet{Agostini2015}. 
It is related to the wall-normal strain of the transverse Stokes layer ($\partial \overline{W}/\partial y$), in particular its phase-wise rate of change referred to as the Stokes strain rate (SSR) $\partial^2\overline{W}/\partial y\partial \xi$, where $\xi$ is the phase.
For a period $T^+ = 200$, \citet{Touber2012} showed a strong attenuation of the near-wall streaks when the SSR was high, while the streaks recovered and were periodically tilted when the Stokes strain was near-constant (referred to as \textit{lingering} in their work) and consequently $\text{SSR}\approx 0$.
These two phases of high SSR and near-zero SSR were classified in \citet{knoop2025response} as \textit{subphase I} and \textit{subphase II}, respectively.

Figure~\ref{fig:intro-recovery} shows results on the streamwise evolution of a drag-reduced TBL over the downstream section of the experimental setup, where the forcing effect had fully developed. 
$C_f$ was determined locally at the center of each belt using a modified composite fit as detailed in appendix A of \citet{knoop2025response}. Its streamwise evolution in figure~\ref{fig:intro-recovery}(a) shows a clear DR effect compared to the non-actuated reference in black, confirms the streamwise-homogeneous response for the sub-optimal (green; $\Lambda_x^+ = 471$) and near-optimal (blue; $\Lambda_x^+ = 942$) forcing regimes, and clearly shows the phase-wise variation for post-optimal conditions (red; $\Lambda_x^+ = 1884$). To demonstrate this recovery effect, the streamwise evolution of the streamwise-normal stresses associated with streaks and the near-wall dynamics are assessed.
In figure~\ref{fig:intro-recovery}(b), the streamwise modulation of the Reynolds streamwise stress, $\overline{uu}^+(x,y) - \langle \overline{uu} \rangle ^+_x (y)$, reveals a periodic alternation of attenuation (blue contours) and recovery (red contours). Here, the $\langle \dots\rangle_x$ operator indicates streamwise averaging over an integer number of $\Lambda_x$.
Using figure~\ref{fig:intro-recovery}(c), the TBL response is related to the $\text{SSR} = \partial^2\overline{W}/\partial y\partial x$ based on the laminar spatial Stokes layer solution for a SqW forcing derived in \citet{Knoop_2024}. 

Owing to the discrete nature of the SqW, \textit{subphase I} is confined to localized high-magnitude impulses where the wall velocity switches direction (refer to red/blue patches that denote $\pm W_w$ imposed by the belts), whereas \textit{subphase II} extends over the rest of the phase where $W_w$ remains constant. 
In agreement with the SSR mechanics proposed in the literature, a significant attenuation of $\overline{uu}$ occurs downstream of \textit{subphase I} and the turbulence then gradually recovers during \textit{subphase II}. 
This intra-phase recovery effect  (i.e., the cycle of periodic turbulence attenuation and recovery) is only apparent in post-optimal forcing conditions where \textit{subphase II} extends for a sufficiently long fetch of $\mathcal{L}^+ \gg 500$ \citep{knoop2025response}.

Due to the difference in response between the mean-flow and turbulent fluctuations, the trends in $C_f$ are out-of-phase with the variation of turbulence statistics. This effect is not the focus of the present study; its underlying mechanism is the development of an internal boundary layer along \textit{subphase II} as elucidated in \citet{knoop2026internal}. The present study instead focuses on characterizing the coherent turbulent structures and their recovery effects for post-optimal forcing conditions.

\subsection{Research approach \label{sec:intro-research approach}}
In this study, we analyze the behavior of drag-producing structures associated with the Reynolds shear stress under the effect of SqW spanwise forcing.
We provide the structural interpretation of the drag-reduction mechanism that remained unaddressed in the statistical analysis of \citet{knoop2025response}. In particular, we aim to elucidate the recovery mechanisms of the post-optimal case.
The dataset of \citet{knoop2025response} was reprocessed with particle tracking velocimetry (PTV) to resolve the instantaneous velocity fields without the spurious noise inherent to correlation-based PIV, required for the current analysis.
We first characterize the global response of the three forcing regimes across the entire actuation surface in \S\ref{sec: global flow organization}.
The instantaneous flow snapshots (\S\ref{subsec:instantaneous_flow}) qualitatively show that forcing suppresses both Q2 and Q4 motions, while the turbulence statistics (\S\ref{subsec:turbulence_statistics}) quantify this global suppression of near-wall motions associated with the Reynolds shear stress.
Quadrant analysis (\S\ref{subsec: Quadrant Analysis}) is used to investigate the effect of wall-forcing on the inter-dependence of the $u$ and $v$ fluctuations.
A novel identification and investigation of the recovery mechanism are performed in the spectral (\S\ref{subsec: Energy Spectra}) and bursting (\S\ref{subsec: Bursting}) analyses.
We then exploit the spatial inhomogeneity of the post-optimal case in \S\ref{sec: Intra-Phase Recovery Effects} to resolve the intra-phase attenuation–recovery cycle by exploring the modulation of the quadrant events (\S\ref{subsec: Spatial Variation of Q-events}) and recovery of small-scale structures using wavelet and bursting analyses (\S\ref{subsec: Wavelet analysis}).

\section{Methodology \label{sec:methods}}
This study builds on the experiments and data analysis reported in \citet{knoop2025response}. 
Here, the key methodology is summarized; for further details, the reader is referred to the original publication.

Experiments on a ZPG TBL were conducted in a subsonic open-return wind tunnel at the Delft University of Technology.
A schematic of the experimental setup is shown in figure~\ref{fig:expSetup}.
The wall forcing setup was installed 2.5\:m downstream of the boundary layer tripping location in a flat-plate TBL test section \citep{dacome2024opposition}.
A local $(x,y,z)$ coordinate system with its origin at the leading edge of the actuation surface is used to present the results herein. 

Experiments were conducted at a free-stream velocity of $U_\infty = 5$\:m/s. At the inflow plane of the actuation surface  (i.e., $x=0$), the TBL characteristics were boundary-layer thickness $\delta_0=70$\:mm, skin-friction velocity $U_{\tau0} = 0.198$\:m/s, corresponding to a friction Reynolds number $Re_\tau \equiv U_{\tau0}\delta_0/\nu= 960$.
Viscous scaling is adopted using $\nu$ and either a reference $U_{\tau0}$ of the non-actuated flow or the actual $U_{\tau}$ of the drag-reduced flow, denoted by superscripts $+$ and $*$, respectively.

\subsection{Wall-forcing setup and dataset\label{sec:ForcingSetup}}

\begin{figure}[t]
    \centering
    \makebox[\textwidth][c]{\includegraphics[width=17cm]{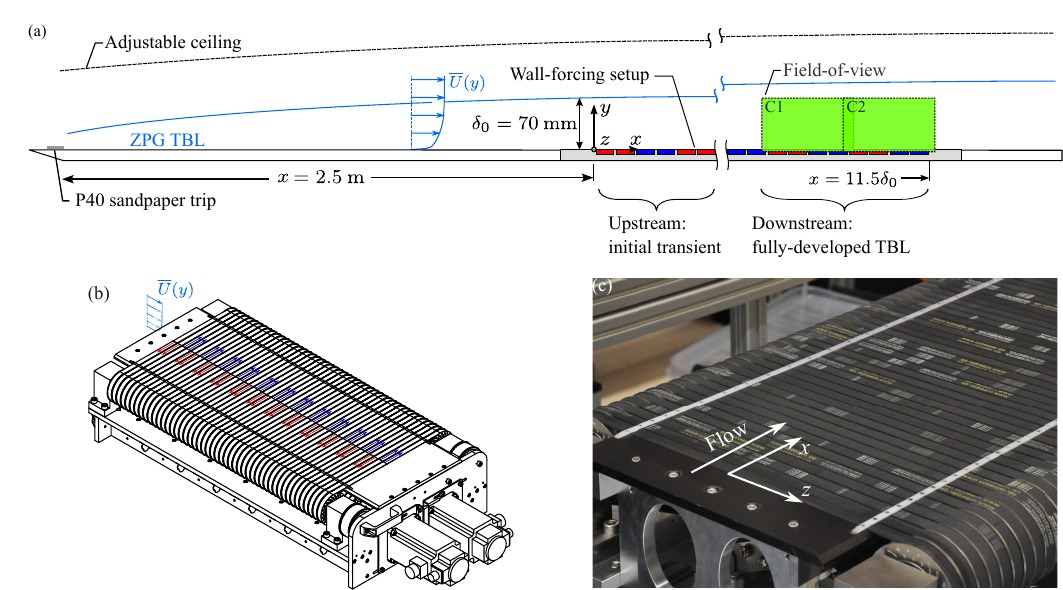}}
    \caption{(a) Schematic of the experimental wall-forcing setup; red/blue patches indicate the steady running belts of the setup and their positive/negative motion direction; the situation shows the near-optimal waveform case with $S=4$. (b) Schematic representation of the wall-forcing setup, and (c) picture of the actuation surface. Reproduced from \citet{knoop2025response}.}
    \label{fig:expSetup}
\end{figure}

The experimental dataset was obtained from \citet{knoop2025response}, with the spanwise wall forcing imposed via a series of spanwise running belts. The belts run at a steady velocity and in opposing spanwise directions to generate a spatially varying SqW spanwise wall velocity, as defined in equation \eqref{W_w}.
The original study already established the initial spatial transient of the forcing effect, which is indicated in figure~\ref{fig:expSetup}(a) at the upstream side of the actuation surface. 
Here, we focus on the fully established forcing effect achieved over the downstream end of the actuation surface.

The wall-forcing setup comprises 48 belts, each with a width (i.e., streamwise fetch) of 15 mm, arranged along the streamwise direction with a 2-mm spacing between belts.
A single periodic \textit{belt element} is therefore $l_x = 2+15 = 17$\:mm.
The actuation surface extends over $816\times 249$\:mm$^2$ or $\approx 11.5\times 4\ \delta_0^2$ in the streamwise and spanwise directions.
The forcing wavelength $\Lambda_x = S\times l_x$ can be varied in a discrete fashion by changing the number of belt elements $S$ that constitute a single waveform, whereas the forcing amplitude is independently controlled by setting the belt velocity.
This independent control over the forcing wavelength $\Lambda_x$ and amplitude $A$ is a unique aspect of this setup since these are parameters typically coupled in experimental implementations \citep{Marusic2021AnReduction, Bird2018ExperimentalWaves, Kempaiah20203-dimensionalOscillation, fumarola2024simultaneous}.

\citet{knoop2025response} originally conducted 2D-2C particle image velocimetry (PIV) in the $(x,y)$ plane. 
As indicated in figure~\ref{fig:expSetup}(a), a two-camera setup captured the forcing effect over the 8 downstream belts, with a corresponding field-of-view of $140\times 49$\:mm$^2$ or $\approx 2 \times 0.7 \ \delta_0^2 $ in the $(x,y)$ directions.
A total of $ N_s = 2000$ uncorrelated snapshots were acquired at 8\:Hz. To investigate coherent flow structures, low-magnitude spurious noise from the original correlation-based PIV was removed by reprocessing the data using particle tracking velocimetry (PTV), employing the \textit{particle tracking from PIV} algorithm in the LaVision DaVis software.
The individual particle tracks ($\approx 500,000$ per snapshot)  were binned in $32\times 32$ pixel bins with a 75\% overlap factor, compared to the $16\times 16$ pixel interrogation windows of the original PIV dataset. Using a larger window led to slightly higher attenuation of the turbulence statistics due to spatial averaging \citep{lee2016validating}, but the results reproduced herein are in overall agreement with \citet{knoop2025response}.
The viscous size of one bin was $\approx 12.8 \nu/U_{\tau0}$, with a vector pitch of $\approx 3.2 \nu/U_{\tau0}$.

The present dataset, summarized in Table~\ref{tab:actuation_conditions}, consists of a non-actuated case, i.e., non-running belts, and three actuated cases with wavelengths $\Lambda_x^+ =[471,942,1884]$ while keeping the forcing amplitude fixed at $A^+ = 12$.
For the three wavelengths, which were realized by increasing the number of belts $S = [2,4,8]$, the streamwise extent of the field-of-view captured 4, 2, and 1 complete wavelengths, respectively. 
The intermediate case of $\Lambda_x^+ \approx 1000$, corresponding to the near-optimal DR regime \citep{Viotti2009StreamwiseReduction, knoop2025response}, is referred to as near-optimal, while the shorter and longer wavelengths are termed sub- and post-optimal, respectively.

\begin{table}[t]
\centering
\caption{Overview of the wall-forcing cases, actuation parameters, and DR reported in \citet{knoop2025response}}
\label{tab:actuation_conditions}
\begin{tabular}{lrrrrrrr}
\hline\hline
Case& $S$ & $\Lambda_x$ (mm) & $A$ (m/s) & $\Lambda_x^+$ & $A^+$ & DR (\%) & Line style \\
\hline
Non-actuated & -- & --  & --   & --   & --   & --   & \legline{black}{{\blacktriangle}} \\
Sub-optimal  & 2  & 34  & 2.46 & 471  & 12.0 & 32.1 & \legline{cb2green}{\blacklozenge} \\
Near-optimal & 4  & 68  & 2.46 & 942  & 12.0 & 38.1 & \legline{cb2blue}{\scalebox{1.8}{$\bullet$}}\\
Post-optimal & 8  & 136 & 2.46 & 1884 & 12.0 & 36.3 & \legline{cb2red}{\blacksquare} \\
\hline\hline
\end{tabular}
\end{table}

\subsection{Scale-by-scale analysis}
To quantify the effect of wall forcing on scale organization in the TBL, wall-normal spectra of the velocity (co-)variances are studied. The velocity fluctuations are Fourier transformed in the streamwise direction, indicated by the $\widehat{\cdot}$ operator: 
\begin{equation}
    \widehat u_i(k_x,y,t) =\int_{-\infty}^{\infty} u_i(x,y,t)e^{-\mathrm{i}k_x x}\mathrm{d}x,
\end{equation}
where $k_x$ is the streamwise wavenumber with $\lambda_x = 2\pi/k_x$ the streamwise wavelength.
The co-spectral density between velocity $u_i$ and $u_j$ is

\begin{equation}
   \phi_{ij}(k_x,y) = \mathrm{Re}\!\left[\,\overline{\widehat{u}_i^{*}(k_x,y,t)\,\widehat{u}_j(k_x,y,t)}\,\right] \big/ \Delta k_x ,
\end{equation}
where $\mathrm{Re}$ denotes the real part, $*$ denotes the complex conjugate, overline denotes time-averaging and $\Delta k_x$ is the wavenumber spacing.
A wavelet transform is used to study the streamwise variation of the scale organization. Instead of the Fourier transform that convolves the signal with a periodic sine/cosine function, the wavelet transform convolves the signal with a scaled mother wavelet that is localized in space. The choice of mother wavelet determines the inherent trade-off between spatial and spectral resolution. In this study, the Morlet (Gabor) wavelet is selected \citep{Ashmead_2012, torrence1998practical}, defined as
\begin{equation}
\psi(\eta)
=
\frac{\pi^{-1/4}}
{C_0}
\left[
\exp\left(\mathrm{i}k_0\eta\right)
-
\exp\left(-\frac{k_0^2}{2}\right)
\right]
\exp\left(-\frac{\eta^2}{2}\right),
\label{eq:morlet}
\end{equation}
with normalization constant
\begin{equation}
    C_0 = \sqrt{
1
- 2\exp\left(-\frac{3k_0^2}{4}\right)
+ \exp\left(-k_0^2\right)
}.
\end{equation}
In equation \eqref{eq:morlet}, $\eta$ is a non-dimensional 'space' parameter and $k_0$ is the non-dimensional central wavenumber that governs the balance between spatial and spectral localization. Increasing $k_0$ effectively lengthens the wavelet, improving scale resolution at the expense of spatial resolution. 
The second exponential term in \eqref{eq:morlet} and normalization constant $C_0$ are included to satisfy the admissibility condition for small values of $k_0$; instead, when $k_0>5$, $\exp(-k_0^2/2) \approx 0$ and $C_0 \approx 1$.
A central wavenumber $k_0$ of 2.5 was chosen to balance spatial localization against wavenumber resolution, ensuring that the spatial scales and their streamwise occurrence could be clearly distinguished. The velocity fluctuations are wavelet transformed in the streamwise direction, indicated by the $\widetilde{\cdot}$ operator:

\begin{equation}
    \widetilde u_i (x, y, t, s(k_x)) = \frac{1}{\sqrt{s}}\int_{-\infty}^{\infty} u_i(x',y,t) \,
    \psi^*\!\left( \frac{x' - x}{s} \right)
    dx'.
    \label{eq:wavelet_transform}
\end{equation}
The wavelet scale is related to the physical streamwise wavenumber $k_x$ through its scaling-factor $s$ expressed in meters \citep{Ashmead_2012, torrence1998practical}:
\begin{equation}
    s = \frac{\left( k_0 + \sqrt{2 + k_0^2} \right)}{2 k_x}.
    \label{eq:s_normalizing}
\end{equation}
Because the velocity $u_i$ is finite, errors will occur at the beginning and end of the wavelet power spectrum. This region, influenced by the edges, is indicated by the cone of influence (COI). Now the wavelet spectrum can be defined as:

\begin{equation}
    \Psi_{ij}(x,y,k_x) = \mathrm{Re}\!\left[\,\overline{\widehat{u}_i^{*}(x,y,t,k_x)\,\widehat{u}_j(x,y,t,k_x)}\,\right] \big/ \Delta k_x .
\end{equation}
\subsection{Bursting analysis \label{sec:method-bursting}}
Bursting events are identified using variable-interval space-averaging (VISA) as introduced in \citet{Kim1985}.
This method detects intermittent bursts of high turbulent kinetic energy production by identifying regions of high local variance of the streamwise velocity fluctuations.
The local variance is computed over a streamwise window of length $L_x$, corresponding to the characteristic length scale of the flow phenomena of interest: 
\begin{equation}
\mathrm{var}(x,y_{ref},t) = \frac{1}{L_x} \int_{x-L_x/2}^{x+L_x/2} \big[ u(x', y_\mathrm{ref}, t) \big]^2 dx'
- \left[ \frac{1}{L_x} \int_{x-L_x/2}^{x+L_x/2}u(x', y_\mathrm{ref}, t)  dx' \right]^2 ,
\end{equation}
where $y_{ref}$ is the reference wall-normal height for detection.

Bursting is identified using a detection function based on the magnitude of the local variance and the sign of the streamwise gradient of the streamwise fluctuations 
\begin{equation}
\label{eq:burst-detection}
D(x,y_\mathrm{ref},t) =
\begin{cases} 
1, & \text{if } \mathrm{var}(x,y_\mathrm{ref}, t) > K \langle \overline{uu}\rangle_x(y_\mathrm{ref}) \text{ and } \frac{\partial u}{\partial x} < 0 \\[1ex]
0, & \text{otherwise}
\end{cases}
\end{equation}
where $K = 1$ is the threshold factor, a commonly adopted value in the literature \citep{CHEN2021108811, hasan2025intrinsic, johansson1991evolution}.
The second criterion, $\partial u/\partial x < 0$, isolates regions of decelerating velocity. Conceptually, these regions are associated with a sweep motion (Q4) leading to and impinging onto an ejection motion (Q2), where a Q2 motion originates from the wall while the Q4 motion originates from farther away from the surface \citep{offen1975bursting}.

A bursting event and its streamwise location $x_b$ are defined as the position of the local maximum of the detection function, obtained using a peak-finding algorithm.
The conditionally averaged fields of a quantity $\chi$, centered around the burst location, were obtained by averaging over the number of detected burst events $N_b$:
\begin{equation}
    \widecheck{\chi}(\Delta x,y)
    =
    \frac{1}{N_b}
    \sum_{n=1}^{N_b}
    \chi(x_b+\Delta x,y,t),
\end{equation}
where the $\widecheck{\cdot}$ sign denotes conditional averaging.
The bursting frequency is defined as the number of detected bursts normalized by the number of PTV snapshots ($N_s = 2000$), i.e., $f_b = N_b/N_s$.

\section{Global flow organization \label{sec: global flow organization}}
This section examines the effect of spatial SqW-type wall forcing on the organization of the turbulent flow and its coherent structures. To establish the global effects, all statistical results (except for \S\ref{subsec:instantaneous_flow}) concern streamwise averaging over an integer number of forcing wavelengths across the actuation surface, which is denoted by $\langle...\rangle_x$. 
The local variation of coherent structures in relation to the turbulence recovery effect under post-optimal forcing conditions is further detailed in \S\ref{sec: Intra-Phase Recovery Effects}.

\subsection{Instantaneous flow \label{subsec:instantaneous_flow}}
\begin{figure}[t]
    \centering
    \includegraphics[width=12.8cm]{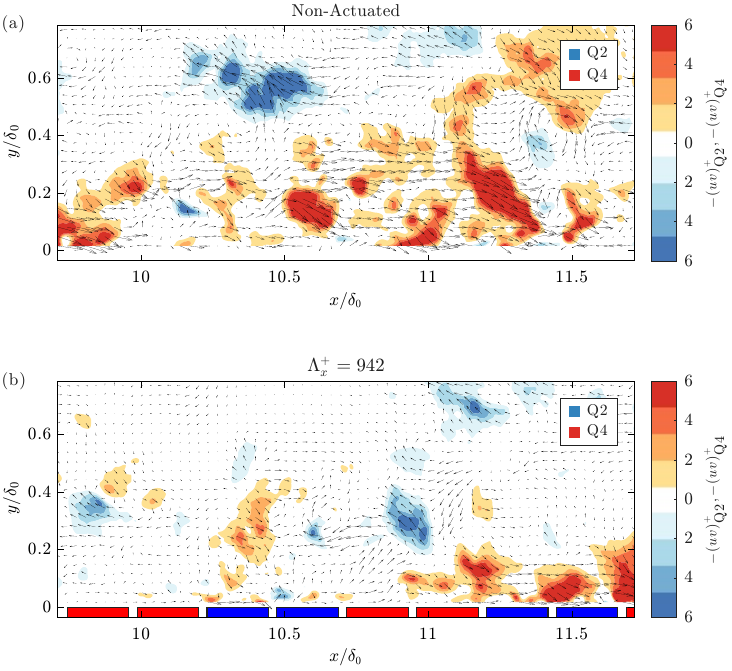}
    \caption{Instantaneous flow for a randomly selected frame: (a) non-actuated flow and (b) near-optimal forcing conditions. Vectors show in-plane $(u,v)$ fluctuations. Contours show the instantaneous Reynolds shear stress $-(uv)_{\text{Q2}}^+$ for ejections in blue and $-(uv)_{\text{Q4}}^+$ for sweeps in red. The colored patches in (b) indicate the belts and the $\pm W_w$ spanwise wall velocity imposed.}
    \label{fig: q2q4ninstflowfield}
\end{figure}

This subsection establishes, in a qualitative sense, the effect of wall-forcing on the instantaneous flow structure.
Figure~\ref{fig: q2q4ninstflowfield} shows instantaneous snapshots of the TBL, comparing (a) the non-actuated case to (b) wall-forcing at near-optimal DR conditions. The colored contours show ejections (Q2) in blue and sweeps (Q4) in red, and the overlaid vectors show the $(u,v)$ velocity fluctuations.
In the non-actuated case, a region of large-scale sweeps extends across almost the entire measurement field-of-view ($9.8 \leq x/\delta_0 \leq 11.6$) and exhibits a clear inclination away from the wall over a large spatial extent of approximately $2\delta_0$. This behavior is consistent with the idea proposed by \citet{Adrian2007HairpinTurbulence} of hairpin vortices and their organization into packets.
The effect of wall-forcing on the structure of the TBL is clearly revealed in (b). A global reduction in velocity fluctuations is evident from the reduced magnitude of the velocity vectors. The colored contours show that ejections and sweeps are strongly suppressed in magnitude, occurrence, and spatial organization by the wall forcing.

\subsection{Turbulence statistics \label{subsec:turbulence_statistics}}
\begin{figure}[t]
    \centering
    \includegraphics[width=12.8cm]{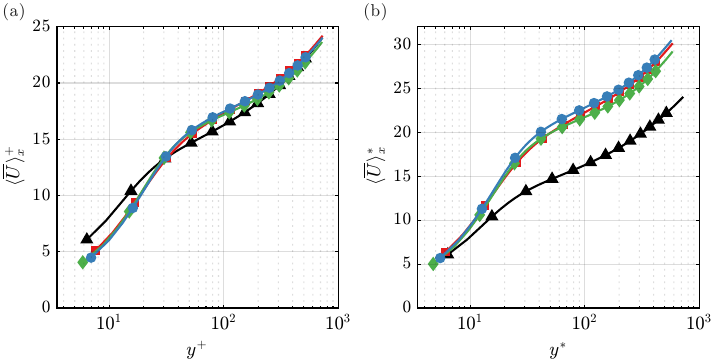}
    \caption{Wall-normal profiles of the mean streamwise velocity. Statistics are normalized by (a) the reference friction velocity $U_{\tau0}$, (b) the actual friction velocity $U_\tau$ of the drag-reduced flow. The linestyles correspond to: \legline{black}{{\blacktriangle}} non-actuated reference, and actuation at \legline{cb2green}{\blacklozenge} sub-optimal ($\Lambda_x^+=471$), \legline{cb2blue}{\scalebox{1.8}{$\bullet$}} near-optimal ($\Lambda_x^+=942$), and \legline{cb2red}{\blacksquare} post-optimal ($\Lambda_x^+=1884$) conditions.}
    \label{fig:mean_velocity}
\end{figure}

The strong turbulence suppression observed instantaneously in \S~\ref{subsec:instantaneous_flow} and the DR effect produced by the wall-forcing are quantitatively supported by the streamwise-averaged turbulence-statistics profiles.
Figure~\ref{fig:mean_velocity} shows the mean streamwise velocity normalized by (a) the reference friction velocity $U_{\tau0}$ of the non-actuated TBL, and (b) the actual $U_\tau$ of the individual actuated cases (i.e., of the drag-reduced skin-friction velocity).
Normalizing by $U_{\tau0}$ in figure~\ref{fig:mean_velocity}(a) highlights absolute changes and, for the actuated cases, exhibits a reduction of near-wall velocity ($y^+<30$). This reduction indicates a thickening of the viscous sublayer and a reduction of the near-wall velocity gradient, and consequently of the skin-friction drag.
Normalizing by the actual friction velocity $U_\tau$ in figure~\ref{fig:mean_velocity}(b) yields a collapse in the viscous sublayer for all cases.
The actuated cases are characterized by an upward shift of the log-layer that is indicative of DR through the log-law shift by $\Delta B$ \citep{Gatti_2016}, and agrees with the DR values reported in Table~\ref{tab:actuation_conditions}.
Figure \ref{fig:mean_velocity}(b) shows that the near-optimal case exhibits the largest DR, followed by the post-optimal, then the sub-optimal case. 

\begin{figure}[t]
    \centering
    \makebox[\textwidth][c]{\includegraphics[width=17cm]{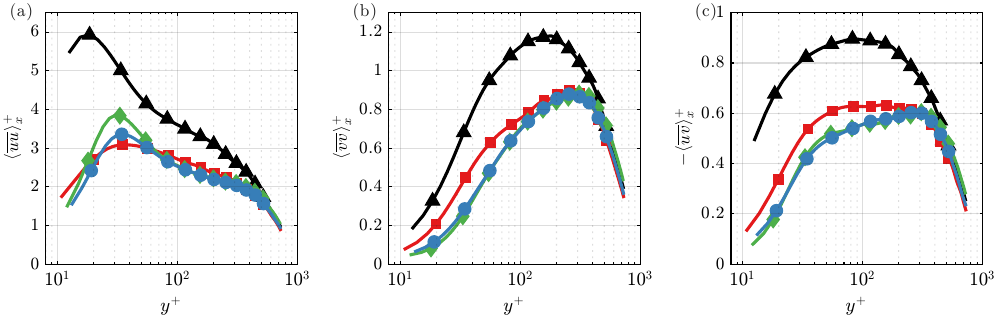}}
    \caption{Wall-normal profiles of second-order statistics, (a) streamwise normal stress, (b) wall-normal normal stress, and (c) Reynolds shear stress. The linestyles correspond to: \legline{black}{{\blacktriangle}} non-actuated reference, and actuation at \legline{cb2green}{\blacklozenge} sub-optimal ($\Lambda_x^+=471$), \legline{cb2blue}{\scalebox{1.8}{$\bullet$}} near-optimal ($\Lambda_x^+=942$), and \legline{cb2red}{\blacksquare} post-optimal ($\Lambda_x^+=1884$) conditions.}
    \label{fig:uu_vv_uv_vort}
\end{figure}

In figure~\ref{fig:uu_vv_uv_vort}(a), the streamwise Reynolds stress $\langle \overline{uu}\rangle_x$\,, which is representative of the near-wall streaks, displays a reduction of the inner-peak and a shift away from the wall, in accordance with prior studies \citep{Viotti2009StreamwiseReduction, Quadrio2009Streamwise-travellingReduction}.
This result indicates that the skin-friction reduction is accompanied by a weakening of the near-wall self-sustaining cycle and a thickening of the viscous sublayer.
The reduction in $\langle\overline{vv}\rangle_x^+$ (figure~\ref{fig:uu_vv_uv_vort}b) signifies a diminished wall-normal momentum mixing and varies coherently with the reduction in $-\langle\overline{uv}\rangle^+_x$ (figure~\ref{fig:uu_vv_uv_vort}c) for all cases.
Between the sub- and near-optimal cases, $\langle\overline{vv}\rangle_x$ and $-\langle\overline{uv}\rangle_x$ show negligible differences in the region $y^+<70$ in contrast to $\langle \overline{uu}\rangle_x$.

The effect of forcing wavelength $\Lambda_x^+$ on the reduction of the $\langle \overline{uu}\rangle_x^+$ inner-peak is in accordance with the trend in DR for the sub- and near-optimal cases (refer to Table~\ref{tab:actuation_conditions}), for which the profiles retain a similar shape to the non-actuated case.
On the other hand, for post-optimal conditions, the inner-peak exhibits a broadening and has a significantly lower magnitude than the near-optimal case despite a near-equal DR for both cases.
In \citet{knoop2025response}, this behavior was ascribed to the turbulence recovery effect introduced in \S\ref{sec:intro-recovery}, leading to a non-uniform streamwise distribution of turbulence intensity and a variation of the $\overline{uu}$ peak height (refer to \S\ref{sec:intro-recovery} figure~\ref{fig:intro-recovery}(b)).
Consequently, streamwise averaging $\overline{uu}^+$ leads to the dissimilar shape of the $\langle\overline{uu}\rangle_x^+$ profile in figure~\ref{fig:uu_vv_uv_vort}(a).
Compared to the other drag-reduced cases, $\langle\overline{vv}\rangle_x^+$ and $-\langle\overline{uv}\rangle_x^+$ are significantly less attenuated for post-optimal forcing in the region of $y^+<300$, which highlights an enhancement of wall-normal momentum transfer compared to the sub- and near-optimal conditions.
The shapes of the $\langle\overline{vv}\rangle_x^+$ and $-\langle\overline{uv}\rangle_x^+$ profiles remain tightly coupled, indicating that modifications in wall-normal mixing are directly mirrored in the Reynolds shear stress.
Moreover, the discrepancy between the $\langle\overline{uu}\rangle_x^+$ and the $\langle\overline{vv}\rangle_x^+$ profiles suggests that the recovery effect has a direct impact on $v$ fluctuations, but a more indirect one on the $u$ fluctuations.

\subsection{Quadrant analysis \label{subsec: Quadrant Analysis}}
\begin{figure}[t]
    \centering
    \makebox[\textwidth][c]{\includegraphics[width=17cm]{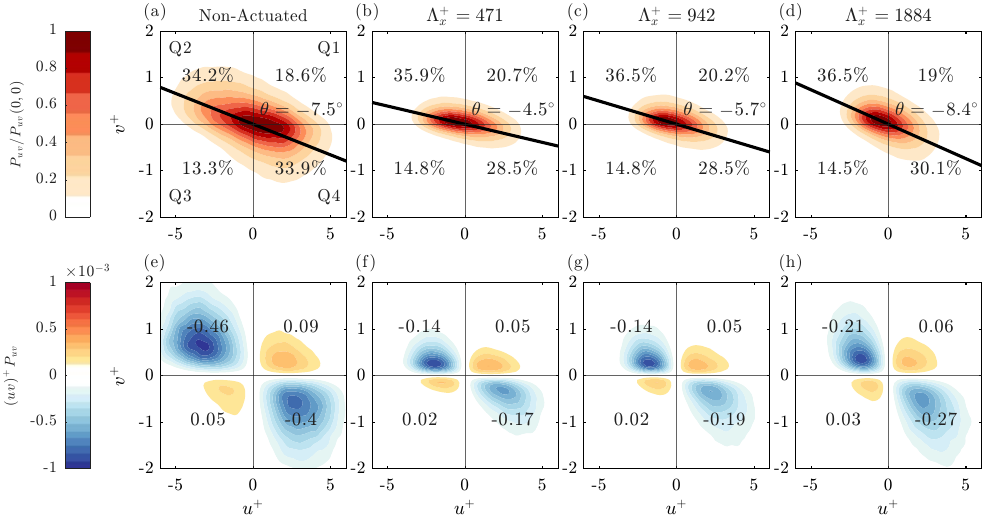}}
    \caption{(a-d) joint probability density function (JPDF) $P_{uv}(u,v)$ normalized by $P_{uv}(0,0)$ and (e-h) $(uv)^+P_{uv}(u,v)$ pre-multiplied JPDF, at $y^+ = 25$. (a,e) Non-actuated reference, and forcing for (b,f) sub-optimal, (c,g) near-optimal and (d,h) post-optimal conditions. Percentages in (a-d) show the probability of occurrence in the respective quadrants, and numbers in (e-h) show the contribution of each quadrant to $\langle uv\rangle_x^+$. The black lines show the major axes of the probability distributions in (a-d) with inclination angle $\theta$.
    }
    \label{fig: Quadrant Analysis}
\end{figure}

\begin{figure}[t]
    \centering
    \includegraphics[width=12.8cm]{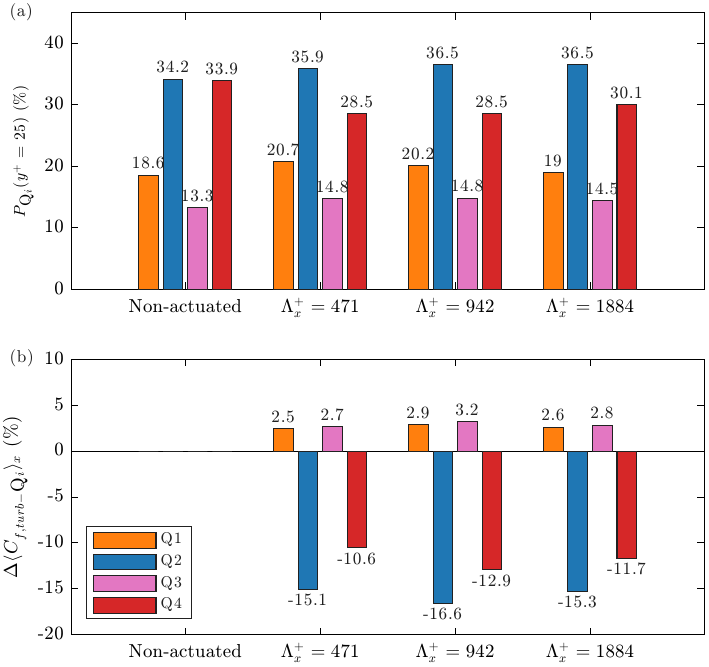}
    \caption{Quadrant analysis at $y^+ = 25$. (a) Percentage probability of each quadrant, (b) percentage change in $C_f$ by each quadrant computed using \eqref{eq:DeltaCf}.}
    \label{fig: Quadrant barchart}
\end{figure}

Quadrant analysis is used to investigate the effect of wall-forcing on the inter-dependence of the $u$ and $v$ fluctuations, which underlies the near-wall coherent structures and their dynamics.
Figures~\ref{fig: Quadrant Analysis}(a-d) show their joint probability distribution $P_{uv}(u,v)$ at $y^+ = 25$, which corresponds to the location where a strong turbulence recovery occurs in the energy spectra (refer to \S\ref{subsec: Energy Spectra}).
The annotated percentages in figures \ref{fig: Quadrant Analysis}(a-d) denote the probability of occurrence of each quadrant event, as visually represented by the bar chart in figure \ref{fig: Quadrant barchart}(a).
When $P_{uv}(u,v)$ is pre-multiplied by $u$ and $v$, shown in figures~\ref{fig: Quadrant Analysis}(e-h), the Reynolds shear stress is recovered by integrating across the function, i.e., $\langle\overline{uv}\rangle_x = \int \int uv\,P_{uv}(u,v)\, \mathrm{d}v\, \mathrm{d}u$ \citep{Wallace2016}, so that the plot directly shows the effect of wall-forcing on the Reynolds shear stress.
The integral contributions of each quadrant to the Reynolds shear stress $\langle\overline{uv}\rangle_{\text{Q}i}$ are indicated by the numbers in figures~\ref{fig: Quadrant Analysis}(e-h).

Significant narrowing of the JPDFs is observed across all actuated cases, indicating a reduced strength of velocity fluctuations, consistent with the instantaneous fields (figure~\ref{fig: q2q4ninstflowfield})  and the second-order statistics profiles (figure~\ref{fig:uu_vv_uv_vort}).
The narrowing of JPDFs is more pronounced along the vertical axis ($v$) than along the horizontal axis ($u$), which reveals that the suppression of $v$ fluctuations is particularly strong.
This anisotropy is quantified by the slope of the JPDF major axis, computed by 
\begin{equation}
    \theta = 0.5\tan^{-1} \left (\frac{2\,\langle\overline{uv}\rangle_x}{\langle\overline{uu}\rangle_x - \langle\overline{vv}\rangle_x}\right),
\end{equation} 
and shown by the black diagonal lines in figure~\ref{fig: Quadrant Analysis}(a-d).
Under actuation, $\lvert \theta \rvert$ drops sharply at sub-optimal forcing and then grows with $\Lambda_x$, ending up slightly exceeding the non-actuated value, consistent with a progressive relaxation of the wall-normal ($v$) suppression.
This enhancement of $v$ fluctuations for the post-optimal case, for which phase-wise attenuation and recovery of turbulence occur (refer to \S\ref{sec:intro-recovery}), suggests that the wall-normal velocity fluctuations are an important aspect of the drag-reduction and turbulence recovery mechanisms.

In the non-actuated case, ejections (Q2) and sweeps (Q4) occur with approximately equal probability, each around 34\%, as displayed in figure~\ref{fig: Quadrant barchart}(a). Under wall forcing, the sweep probability drops markedly to 29--30\%, compensated by a slight increase across the remaining quadrants, yielding a strong Q2–Q4 asymmetry.
This shift is qualitatively observed in figures~\ref{fig: Quadrant Analysis}(a-d), where the probability peak shifts along the major axis from $u>0$ for the non-actuated flow to $u<0$ under wall-forcing, i.e., shifting away from the Q4 events.
Despite near-equal probability of occurrence for Q2 and Q4 in the non-actuated case, they contribute unequally to the Reynolds shear stress (figure \ref{fig: Quadrant Analysis}(e)): $-\langle \overline{uv}\rangle^+_\text{Q2}=0.46$ versus $-\langle \overline{uv}\rangle^+_\text{Q4}=0.40$, indicating that ejections are more intense than sweeps.
Based on figures \ref{fig: Quadrant Analysis}(f-h), wall forcing suppresses ejections more strongly than sweeps, as $-\langle \overline{uv}\rangle^+_\text{Q2}<-\langle \overline{uv}\rangle^+_\text{Q4}$, reversing the trend from the non-actuated flow where the ejections dominate the sweeps (Q2$>$Q4). Notably, this reversal of intensities occurs despite ejections becoming more probable.

These local observations (at $y^+=25$) are reinforced by establishing the global effects of the modulated quadrant motions on the skin friction.
The effect of turbulence in each quadrant on $C_f$ is quantified using $C_{f,turb}$ defined in equation \eqref{eq: Cf wall-normal integration}, which relies on wall-normal integration of the Reynolds shear stress per \citet{Elnahhas2022}. 

We compute the change in each quadrant ($C_{f,turb-\text{Q}i}$) as a percentage of the total non-actuated case $C_{f0,turb}$:
\begin{equation}
    \Delta C_{f,turb-\text{Q}i}(\%) = \frac{C_{f,turb-\text{Q}i} - C_{f0,turb-\text{Q}i}}{C_{f0,turb}} \times 100.
\end{equation}
The present dataset does not permit integration across the full TBL height, as it only extends to $ y = 0.79\delta_0$ ($y^+ = 730$), as shown, for example, in figure~\ref{fig:mean_velocity}. Based on the observation that all $-\langle \overline{uv} \rangle _x$ profiles are invariant for $y^+ > 500$, it is deemed plausible that the energy distribution does not deviate from the non-actuated case beyond the integration bounds. Using a reference $l(x)$ across all cases, the streamwise-averaged $\Delta C_{f,turb-\text{Q}i}$ is accordingly approximated by

\begin{equation}
\label{eq:DeltaCf}
\Delta \langle C_{f,turb-\text{Q}i}\rangle_x(\%) \approx
\frac{\int_0^{0.79\delta_0}
      \left( \langle \overline{uv}\rangle_{\text{Q}i}
           - \langle \overline{uv}\rangle_{0,\text{Q}i} \right) \mathrm{d}y}
     {\int_0^{0.79\delta_0}
      \langle \overline{uv}\rangle_{0}\, \mathrm{d}y}
\times 100 .
\end{equation}
Figure~\ref{fig: Quadrant barchart}(b) confirms that, regardless of a higher occurrence probability, ejection motions constitute the largest reduction in turbulent friction of $\Delta \langle C_{f,turb-\text{Q}2}\rangle_x \approx 15-16\%$, whereas sweeps constitute  $\Delta \langle C_{f,turb-\text{Q}4}\rangle_x \approx 11-13\%$. 
This result, namely that modifications in Q2 cause the largest DR effect, is consistent with  \citet{Yakeno2014} and \citet{Gallorini2022}, and extends their observations on OW and STW forcing to spatial forcing for the first time.

\begin{figure}[t]
    \centering
    \makebox[\textwidth][c]{\includegraphics[width=17cm]{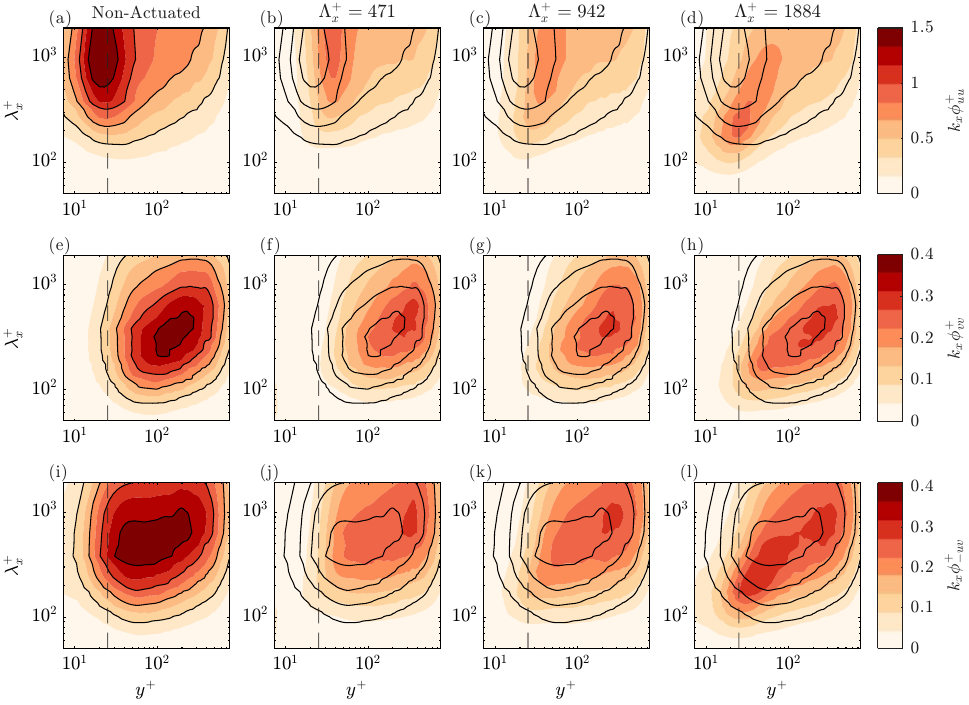}}
    \caption{Pre-multiplied streamwise co-spectra $k_x\phi_{ij}(\lambda_x,y)$ as a function of streamwise wavelength $\lambda_x^+$ and wall-normal distance $y^+$, (a,e,i) for the non-actuated reference and actuation at (b,f,j) sub-optimal, (c,g,k) near-optimal, and (d,h,l) post-optimal conditions. 
    (a-d) Streamwise normal stress, (e-h) wall-normal normal stress, and (i-l) Reynolds shear stress spectra are shown. Black contour lines of the non-actuated case are overlaid on the actuated cases for reference. Vertical dashed lines indicate $y^+=25$.}
    \label{fig:PSDgrid}
\end{figure}

\subsection{Energy spectra \label{subsec: Energy Spectra}}
Figure \ref{fig:PSDgrid} shows the wall-normal pre-multiplied spatial co-spectra of the Reynolds stress components discussed in Figure~\ref{fig:uu_vv_uv_vort}. The black contour lines of the non-actuated case are overlaid in the other cases for reference. The columns, from left to right, correspond to the four test cases, and the rows, from top to bottom, correspond to the three Reynolds stress components.
For the non-actuated case, the locations of the spectral peaks are consistent with the available literature on TBL flow over a flat plate \citep{mathis2009large,baars2024reynolds}.
A broadband energy attenuation is displayed for wall forcing at sub- and near-optimal conditions.
The response is similar for both cases, where the spectral peaks shift away from the wall and towards higher streamwise scales $\lambda_x^+$.

The response for post-optimal conditions is different.
The spectral peak of the streamwise fluctuations in figure~\ref{fig:PSDgrid}(d), which is nearly vertical for $\Lambda_x^+ = 471$ and $942$ (figures~\ref{fig:PSDgrid}b,c), broadens, shifts to smaller scales and toward the wall, and becomes inclined across $10 \lesssim y^+ \lesssim 90$ and $100 \lesssim \lambda_x^+ \lesssim 1000$.
A similar trend is observed for $k_x\phi_{vv}^+$ in figure~\ref{fig:PSDgrid}(h) and the effect is particularly strong for the Reynolds shear stress co-spectrum $k_x\phi_{-uv}^+$ in figure~\ref{fig:PSDgrid}(l).
The enhancement of smaller scales is especially interesting; comparing the post-optimal case to the non-actuated reference in figure~\ref{fig:PSDgrid}(a,e,i) confirms that they are considerably smaller than those typically present in canonical turbulence.
In this work, we refer to those scales that are not energetic in the non-actuated case, but markedly energized in the post-optimal one, as \textit{very small scales}.
A link between the turbulence recovery process under post-optimal forcing and the enhancement of very small scales would appear likely.
This phenomenon is a key finding that, to the best of the authors' knowledge, has not been previously reported for TBL flow subject to spanwise wall forcing.

\begin{figure}[t]
    \centering
    \makebox[\textwidth][c]{\includegraphics[width=17cm]{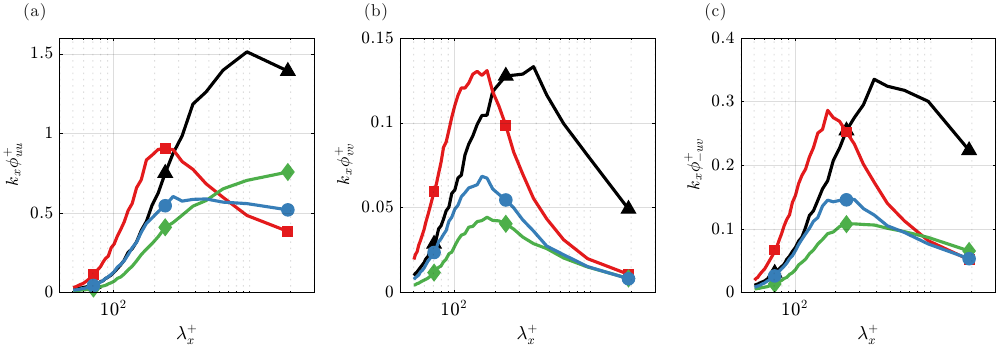}}
    \caption{Pre-multiplied energy spectra at $y^+ = 25$, (a) streamwise normal stress, (b) wall-normal stress, and (c) Reynolds shear stress. The linestyles correspond to: \legline{black}{{\blacktriangle}} non-actuated reference, and actuation at \legline{cb2green}{\blacklozenge} sub-optimal ($\Lambda_x^+=471$), \legline{cb2blue}{\scalebox{1.8}{$\bullet$}} near-optimal ($\Lambda_x^+=942$), and \legline{cb2red}{\blacksquare} post-optimal ($\Lambda_x^+=1884$) conditions.}
    \label{fig:Linespectra}
\end{figure}

Figure~\ref{fig:Linespectra} presents one-dimensional pre-multiplied spectra extracted at $y^+ = 25$, indicated by the vertical dashed lines in figure~\ref{fig:PSDgrid}, which corresponds to the location of the largest energy enhancement of the post-optimal case relative to the non-actuated one ($\Delta\phi_{ij}$ spectra are not shown).
The small-scale energy enhancement at $\lambda_x^+ \lesssim 300$ for the post-optimal case is clearly visible across all spectra.
While the small-scale energy of the streamwise fluctuations (figure~\ref{fig:Linespectra}a) is enhanced, the large-scale energy ($\lambda_x > 800$) representative of low-speed streaks is attenuated more strongly in the post-optimal case than in any other actuated case.
Moreover, the large-scale energy decreases monotonically with $\Lambda_x^+$.
The trend may be favorable for the large-scale forcing strategy proposed in \citet{Marusic2021AnReduction}.
In the post-optimal case, the wall-normal variance spectrum in figure~\ref{fig:Linespectra}(b) shows the strongest relative increase with respect to the near-optimal case, revealing the importance of wall-normal velocity fluctuations in the recovery process.
Taken together, these observations suggest that turbulence recovery originates at very small scales, while the formation of large-scale structures is further suppressed as the forcing wavelength increases.

\subsection{Bursting \label{subsec: Bursting}}
\begin{figure}[t]
    \centering
    \makebox[\textwidth][c]{\includegraphics[width=17cm]{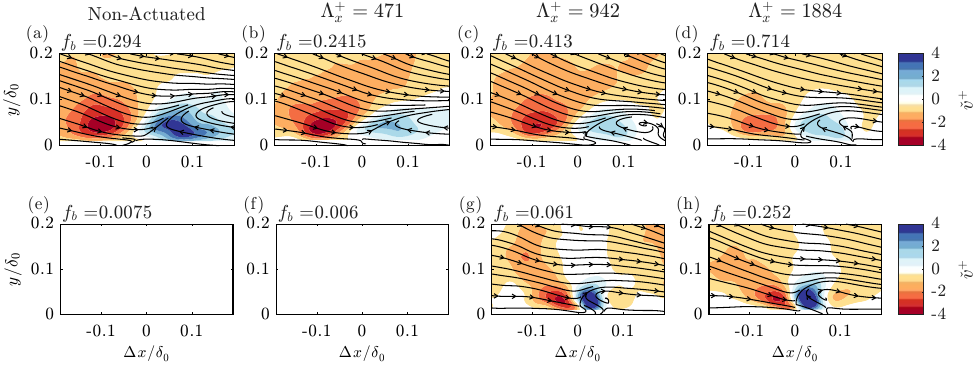}}
    \caption{Wall-normal velocity $\widecheck v^+$ conditionally averaged on decelerating burst events. Streamlines show $(\widecheck u, \widecheck v)$. (a-d) large-scale bursts detected using $L_x^+ = 200$, and (e-h) small-scale bursts detected using $L_x^+ = 50$. From left to right, (a,e) non-actuated reference, and forcing conditions (b,f) sub-optimal, (c,g) near-optimal, and (d,h) post-optimal. Statistically insignificant cases (e,f) with $f_b<0.025$ are not shown.}
    \label{fig: Bursting events}
\end{figure}

This small-scale recovery phenomenon is elucidated using conditionally averaged flow fields on bursts of high TKE production.
VISA analysis (refer to \S\ref{sec:method-bursting}) on large-scale bursting is well documented in the literature \citep{Kim1985, johansson1991evolution, hasan2025intrinsic}, for which a characteristic bursting length scale of order $L_x^+ = 200-250$ is typically considered.
Here, VISA analysis is extended to small-scale bursts, motivated by our observations in \S\ref{subsec: Energy Spectra}. The large- and small-scale length scales are chosen as $L_x^+ = 200$ and $50$, respectively.
The threshold for the burst detection criterion (refer to equation \eqref{eq:burst-detection}) was based on the local $\langle\overline{uu}\rangle_x(y^+ = 25)$ of the individual cases to account for the reduction in turbulent fluctuations by the wall forcing.
The wall-normal velocity conditionally averaged on bursting events ($\widecheck{v}$) is shown in figure~\ref{fig: Bursting events}, and is of particular interest as this component is enhanced the most by post-optimal forcing (refer to figures~\ref{fig:uu_vv_uv_vort} and \ref{fig: Quadrant Analysis}).
We recall that $\widecheck v>0$ corresponds to Q2 ejections, while $\widecheck v < 0$ corresponds to Q4 sweeps.

The top row in figure~\ref{fig: Bursting events} shows $\widecheck v$ conditioned on large-scale bursts.
For the non-actuated case in figure~\ref{fig: Bursting events}(a), the typical structure of a decelerating burst is observed: a sweep motion at $\Delta x<0$ originates from higher up in the TBL and impinges onto an ejection at $\Delta x>0$ that emanates from the wall.
Across all levels of actuation, in figures~\ref{fig: Bursting events}(b-d), the strongest attenuation is observed in Q2 events (in blue).
The increased bursting frequency is attributable to the detection threshold based on $\langle \overline{uu}\rangle_x$ of the individual cases.
Furthermore, the Q4 sweeps (in red) progressively decrease in strength as the forcing wavelength increases.
We conjecture that this is due to the thickening of the Stokes layer.
Analogous to the problem of an impulsively started plate, the local viscous layer would thicken in proportion to the square root of the streamwise extent of constant $W_w$ due to viscous diffusion  (i.e., $\delta_s \propto \sqrt{\Lambda_x}$).
Under this hypothesis, in proportion to $\Lambda_x$, the Stokes layer extends its influence further into the boundary layer, thereby more effectively suppressing the large-scale, outer-layer Q4 motions.

The effect of wall-forcing on small-scale bursting is shown in figures~\ref{fig: Bursting events}(e-h).
In the non-actuated and sub-optimal cases, the results were not statistically significant, as practically no bursts occurred; a threshold on the bursting frequency was set to $f_b = 0.025$ (50 detected instances).
This low statistical occurrence is arguably the reason that small-scale bursting has not been considered in the literature.
The near- and post-optimal cases display a self-similar structure of high-intensity small-scale bursts.
Their bursting frequency increases by a factor of four for the post-optimal case, consistent with the recovery of the $v$ fluctuations with increasing $\Lambda_x$ (refer to figures~\ref{fig: Quadrant Analysis}b-d).
The self-similar structure of these small-scale bursts, as exemplified in figures~\ref{fig: Bursting events}(g,h), hints at a universality in the recovery phenomenon itself.
These results provide strong evidence that the recovery process, which intensifies with the forcing wavelength $\Lambda_x$, enhances small-scale bursts at the wall and acts particularly strongly on the wall-normal fluctuations, while simultaneously suppressing the influence of the outer-layer sweeps.

\section{Intra-phase variation of coherent structures and their recovery \label{sec: Intra-Phase Recovery Effects}}
Whereas \S\ref{sec: global flow organization} established the global flow organization by considering the streamwise-averaged effects, here we address the streamwise inhomogeneity of the flow structure, i.e., its variation along the (spatial) phase of the actuation cycle. Focus is on the post-optimal forcing condition ($\Lambda_x^+ \gg 1000$) that shows a significant streamwise inhomogeneity and turbulence recovery across the phase (i.e., intra-phase) as reported in \citet{knoop2025response} and reviewed in \S\ref{sec:intro-recovery} (see also figure~\ref{fig:intro-recovery}). In this section, we refer to the streamwise phases of high SSR  and near-zero SSR as \textit{subphase I} (where the wall velocity switches sign) and \textit{subphase II} (where the wall velocity is constant), respectively.

\subsection{Streamwise modulation of quadrant events \label{subsec: Spatial Variation of Q-events}}
\begin{figure}[t]
    \centering
    \includegraphics[width=12.8cm]{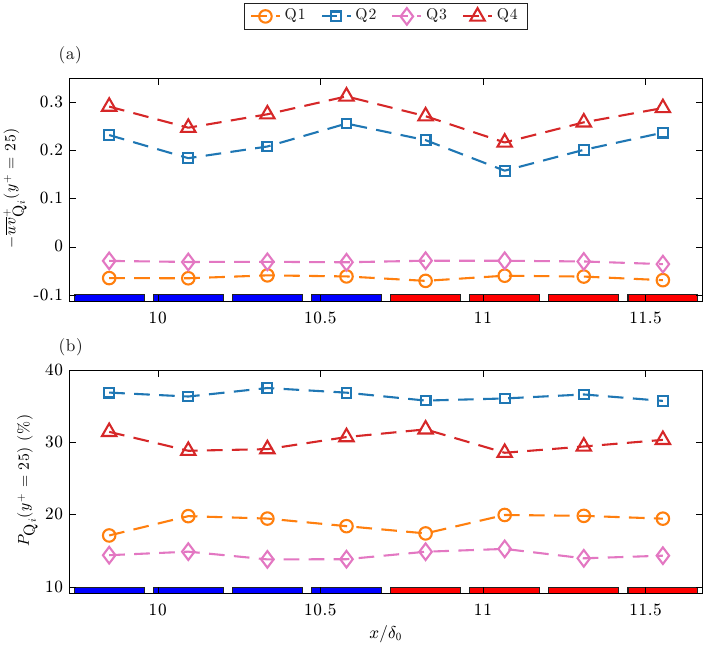}
    \caption{Streamwise analysis of $\Lambda_x^+=1884$ actuation. (a) Variation of quadrant contributions at $y^+ = 25$ to Reynolds shear stress, (b) Percentage occurrences of quadrant events.}
    \label{fig: streamwise distribution of Q-events}
\end{figure}

We examine the streamwise evolution of the quadrant analysis for the post-optimal actuation case.
Recall from \S\ref{sec:intro-recovery} that for this case, i) the turbulence is attenuated downstream of \textit{subphase I} where the wall velocity $W_w$ changes sign at $x/\delta_0 \approx 9.7,\, 10.7$, and ii) turbulence recovery occurs along the rest of \textit{subphase II} where $W_w$ is constant. 
In figure~\ref{fig: streamwise distribution of Q-events}, the sign of $W_w$ is indicated by the red/blue patches.
Figure~\ref{fig: streamwise distribution of Q-events}(a) displays the integral contributions of each quadrant $-\overline{uv}_{\text{Q}i}^+$ to $-\overline{uv}^+$, which are obtained by integrating the pre-multiplied JPDF over a streamwise extent of one belt element of length $l_x$.
No significant variation is observed for Q1 and Q3 while the streamwise modulation of $-\overline{uv}_{\text{Q2}}^+$ and $-\overline{uv}_{\text{Q4}}^+$ is in agreement with the cyclic attenuation/recovery of $-\overline{uv}^+$ reported in \citet{knoop2025response}.

For the probability of occurrence depicted in figure~\ref{fig: streamwise distribution of Q-events}(b), Q1 and Q4 display a notable spatial modulation, whereas the effects on Q2 and Q3 are less pronounced.
There exists a strong coupling with an opposing trend between $P_{\text{Q}1}$ and $P_{\text{Q}4}$, i.e., an increase (decrease) in $P_{\text{Q}1}$ leads to a decrease (increase) in $P_{\text{Q}4}$. 
This correlation can be quantified by computing the covariance, yielding $\text{cov}(P_{\text{Q1}}, P_{\text{Q4}})=-1.13$, which is considerably stronger than $\text{cov}(P_{\text{Q2}}, P_{\text{Q3}})=-0.20$.
This Q1–Q4 coupling shows that, for positive $u$ fluctuations, the attenuation phase is associated with a preference for positive wall-normal fluctuations ($v>0$), while the recovery phase favors negative wall-normal fluctuations ($v<0$). Together, these trends strengthen the evidence that wall-normal velocity fluctuations play a central role in the recovery process.
Why this coupling is far weaker for $u<0$ (Q2/Q3) remains elusive and is a question for future investigation.

\subsection{Spatial recovery of small-scale structures \label{subsec: Wavelet analysis}}
\begin{figure}[t]
    \centering
    \makebox[\textwidth][c]{\includegraphics[width=17cm]{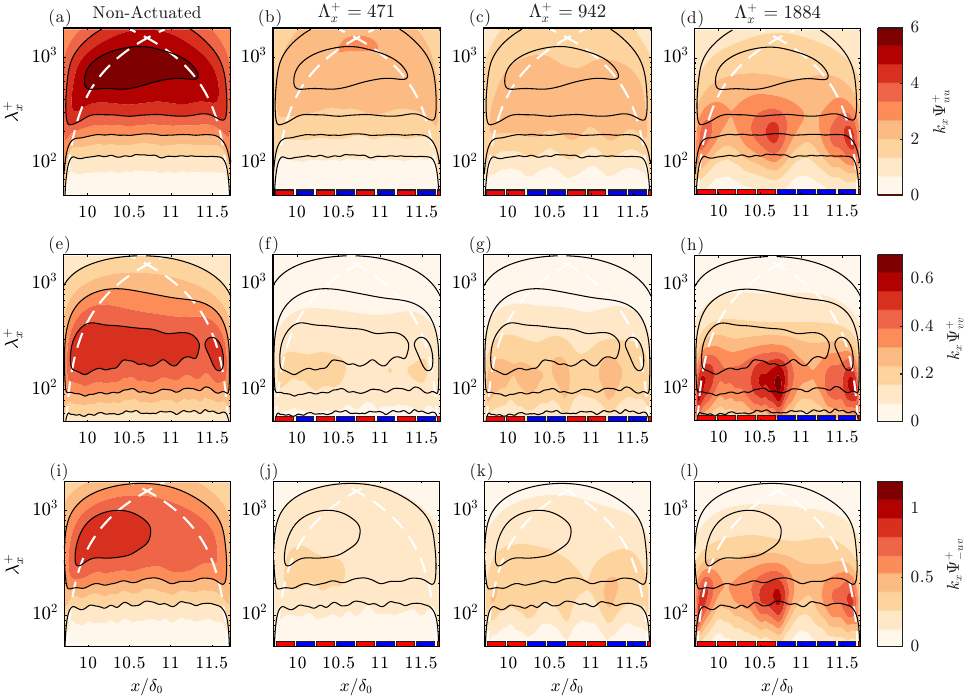}}
    \caption{Pre-multiplied wavelet spectra of (a,e,i) the non-actuated reference, (b,f,j) sub-optimal, (c,g,k) near-optimal, and (d,h,l) post-optimal forcing. Spectra of (a-d) streamwise normal stress, (e-h) wall-normal normal stress, and (i-l) Reynolds shear stress. Actuated cases have belts showing $x$-location of reversals and contours of non-actuated spectra.}
    \label{fig: waveletspectra}
\end{figure}

To establish the spatial variation of the scale organization in relation to the recovery effects, particularly the small-scale enhancement under post-optimal forcing noted in \S\ref{subsec: Energy Spectra}, a wavelet analysis is performed.
The pre-multiplied wavelet co-spectra at $y^+=25$ are shown in figure~\ref{fig: waveletspectra}; the horizontal axis corresponds to the streamwise extent of the flow-field measurements, and the vertical axis corresponds to the local $\lambda_x^+$ of the turbulent structures.
For the non-actuated case in figures~\ref{fig: waveletspectra}(a,e,i), the energy distribution across $\lambda_x^+$ agrees with the line spectra in figure~\ref{fig:Linespectra}.
Outside the cone of influence, plotted in white lines, $\Psi_{ij}^+$ is impacted by edge effects due to the finite wavelet size, as shown in the energy drop-off for $\lambda_x^+ > 500$; nevertheless, the small scales we are interested in are unaffected.
For the sub-optimal case in figures~\ref{fig: waveletspectra}(b,f,j), there is no clear streamwise periodicity in the spatial length scales, whereas a slight periodic variation aligns with the forcing wavelength in the near-optimal case (figures~\ref{fig: waveletspectra}(c,g,k)).

A strong streamwise variation instead appears under post-optimal forcing conditions in figures~\ref{fig: waveletspectra}(d,h,l).
Using the Reynolds shear stress as an example (consistent behavior across all the Reynolds stresses): $k_x\Psi^+_{-uv}$ is shifted to considerably smaller wavelengths, i.e., $\lambda_x^+ \approx 70-300$ compared to $\lambda_x^+ \approx 200-1000$ for the non-actuated case.
In good agreement with the turbulence attenuation/recovery cycle in the integral statistics, $k_x\Psi^+_{-uv}$ is first suppressed at \textit{subphase I}, followed by an enhancement at the very small scales in regions of \textit{subphase II} that is maximum at the half-phase location.
This recovery effect only affects the smaller scales, while $k_x\Psi^+_{-uv}$ at $\lambda_x^+>300$ remains largely unchanged.

\begin{figure}[t]
    \centering
    \makebox[\textwidth][c]{\includegraphics[width=12.8cm]{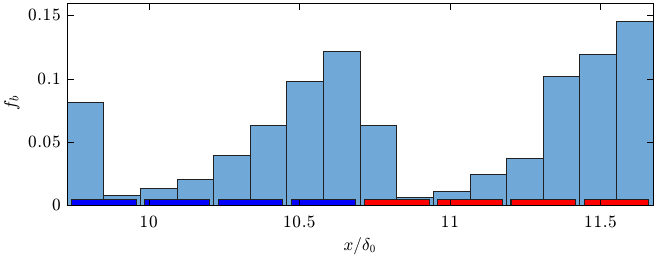}}
    \caption{Streamwise variation of small-scale burst ($L_x^+ = 50$) frequency under post-optimal forcing conditions ($\Lambda_x^+ = 1884$).}
    \label{fig: spatial bursting}
\end{figure}

In \S\ref{subsec: Bursting}, we showed that the recovery process is strongly linked to the enhancement of small-scale bursting events.
To further support our analysis, figure~\ref{fig: spatial bursting} shows the streamwise variation of the small-scale bursting frequency for the post-optimal case.
Consistent with the wavelet co-spectra, the burst frequency drops sharply downstream of each location of sign-reversal of wall velocity $W_w$ (i.e., \textit{subphase I}), after which it gradually recovers in regions of \textit{subphase II} until the next reversal.
Overall, these results clarify the intra-phase recovery mechanism from a coherent-structures perspective, extending the previous statistical characterization in \citet{knoop2025response}.

\section{Summary and conclusion \label{sec: conc}}
This study characterized the coherent structures in a drag-reduced TBL under steady SqW spanwise wall forcing. An innovative aspect of this work is its particular focus on the recovery process, an analysis enabled by the streamwise inhomogeneity of the post-optimal forcing case.
In doing so, we provide the structural interpretation of the DR and turbulence recovery mechanisms, supplementing the statistical analysis in \citet{knoop2025response}.

Across all actuated cases, the SqW forcing globally suppresses the near-wall motions associated with the Reynolds shear stress, as evident throughout our statistical analysis.
Ejections (Q2) become more probable under forcing, yet they undergo the largest reduction in intensity, and therefore their modulation constitutes the largest DR effect (\S\ref{subsec: Quadrant Analysis}).
Our results are consistent with the literature on temporal and spatio-temporal forcing \citep{Yakeno2014, Gallorini2022}, extending these observations to the spatial forcing case for the first time.
The tilting of the major axes of $uv$ joint probability density functions (quadrant analysis) shows that forcing suppresses $v$ fluctuations more strongly, underscoring the key role of wall-normal momentum transport in DR mechanisms.

This asymmetric suppression between the $u$ and $v$ fluctuations, however, weakens as $\Lambda_x$ increases, demonstrating that the $v$ fluctuations recover faster than the $u$ fluctuations, highlighting their direct role in the recovery process that occurs for large $\Lambda_x^+$.
The spatial modulation of coherent structures (\S\ref{sec: Intra-Phase Recovery Effects}) confirms these cyclic attenuation/recovery processes: turbulence is attenuated downstream of each sign-switch of the wall velocity $W_w$ (high SSR - \textit{subphase I}) and recovers where $W_w$ is constant (near-zero SSR - \textit{subphase II}).
A novel key finding of the study is that this recovery phenomenon corresponds to an enhancement of \textit{very small scales}, i.e., $\lambda^+_x \approx$~70--300 (\S\ref{subsec: Energy Spectra}), that are substantially smaller than those in canonical TBLs. 
This enhancement of very small scales is caused by the emergence of small-scale bursting events (\S\ref{subsec: Bursting}). Small-scale bursts are practically non-existent in the non-actuated TBL and the sub-optimal forcing case, but occur frequently in the post-optimal case.
Despite the enhancement of small-scale structures, the larger scales comparable to the near-wall streaks are further suppressed as $\Lambda_x$ increases, highlighting the scale-selective nature of the recovery process.
These results therefore highlight possibilities for forcing at large length/time scales as a drag-reduction strategy, such as in \citet{Marusic2021AnReduction}.
We conjecture that the effect is caused by the thickening of the Stokes layer in proportion to forcing wavelength as $\delta_s \propto \sqrt{\Lambda_x}$, causing a deeper penetration into the TBL and a more effective interaction with the larger-scale Q4 motions away from the wall (\S\ref{subsec: Bursting}).

Taken together, our analysis elucidates the mechanism of intra-phase recovery from a structural perspective, contributing to a fundamental understanding of near-wall turbulent dynamics. 
However, several questions remain open and motivate future work.
Why the Q1–Q4 coupling occurs predominantly for $u > 0$ (\S\ref{subsec: Spatial Variation of Q-events}) requires further investigation.
The near- and post-optimal cases exhibit a self-similar small-scale burst structure, suggesting a possible universality of the recovery phenomenon that should be explored in future studies.

\begin{acknowledgments}
This work was initiated in the context of the second-year TU Delft Aerospace Engineering course Test, Analysis and Simulation (AE2224-1) in the 2024--2025 academic year, and subsequently developed into the present scientific paper. The authors thank Eva Rodrigues, Marton Meszaros, Ian Holmes, Marcos Merino Francos, and Jasper van Ruiten for their contributions to the preliminary work that formed the basis of this study. 
Max W. Knoop and Bas W. van Oudheusden sincerely thank Dr. Ir. Gerrit E. Elsinga for his suggestion to explore the flow structure associated with turbulent frictional drag and the fruitful discussions we had thereafter together with Dr. Ir. Ferry F.J. Schrijer. 
The work is financially supported by the Netherlands Enterprise Agency, under grant number TSH21002.

\end{acknowledgments}

\section*{Author contributions}
T.B.: Conceptualization, Formal Analysis, Writing -- Original Draft, Writing -- Review \& Editing.
V.A.K.: Conceptualization, Formal Analysis, Writing -- Original Draft, Writing -- Review \& Editing.
P.M.: Visualization, Writing -- Review \& Editing.
R.A.A.K.: Formal Analysis, Writing -- Original Draft, Writing -- Review \& Editing.
M.W.K.: Conceptualization, Formal Analysis, Investigation, Methodology,  Data Curation, Writing -- Original Draft, Writing -- Review \& Editing, Supervision.
B.W.v.O.: Writing -- Review \& Editing, Supervision.

\section*{Data Availability}
A full dataset accompanying the initial publication \citep{knoop2025response} is available \href{http://www.doi.org/10.4121/adb0b003-4912-4bdb-b12f-e56d18c3b937}{online} \citep{Dataset}. The present data will be made available upon reasonable request to the corresponding author. 

\bibliography{references}

\end{document}